\documentclass{article}

\usepackage{arxiv}

\usepackage[utf8]{inputenc} 
\usepackage[T1]{fontenc}    
\usepackage{hyperref}       
\usepackage{url}            
\usepackage{booktabs}       
\usepackage{amsfonts}       
\usepackage{nicefrac}       
\usepackage{microtype}      
\usepackage{lipsum}		
\usepackage{graphicx}
\usepackage[numbers]{natbib}
\usepackage{doi}

\title{Inverted and programmable Poynting effects in metamaterials}

\author{{Aref Ghorbani\textsuperscript{1*}, David Dykstra\textsuperscript{2}, Corentin Coulais\textsuperscript{2}, Daniel Bonn\textsuperscript{2}, Erik van der Linden\textsuperscript{1}, Mehdi Habibi\textsuperscript{1$\dagger$}} 
	\vspace{12pt} \\ 
	\textsuperscript{1} Laboratory of Physics and Physical Chemistry of Foods, Wageningen University, 6708WG Wageningen, The Netherlands\\
	\textsuperscript{2} Institute of Physics, University of Amsterdam, 1098XH Amsterdam, The Netherlands
	\vspace{12pt} \\
}

\renewcommand{\shorttitle}{Inverted and programmable Poynting effects}

\hypersetup{
pdftitle={A preprint for arxiv},
pdfsubject={cond-mat.soft},
pdfauthor={Aref Ghorbani},
pdfkeywords={Poynting effect, Inverted Poynting effect, metamaterials},
}

\begin{document}
\maketitle

\begin{abstract}
			The Poynting effect generically manifests itself as the extension of the material in the direction perpendicular to an applied shear deformation (torsion) and is a material parameter hard to design. Unlike isotropic solids, in  designed structures, peculiar couplings between shear and normal deformations can be achieved  and exploited for practical applications. Here, we engineer a metamaterial that can be programmed to contract or extend under torsion and  undergo nonlinear twist under compression. First, we show that our system exhibits a novel type of inverted Poynting effect, where axial compression induces a nonlinear torsion. Then the Poynting modulus of the structure is programmed from initial negative values to zero and positive values via a pre-compression applied prior to torsion. Our work opens avenues for programming nonlinear elastic  moduli of materials and tuning the couplings between shear and normal responses  by rational design.  Obtaining inverted and programmable Poynting effects in metamaterials inspires diverse applications from designing machine materials, soft robots and actuators to engineering biological tissues, implants and prosthetic devices functioning under compression and torsion.
\end{abstract}



\par \textit{Introduction:}
The Poynting effect is a surprising non-linear elastic effect that makes, in the original experiment of Poynting \cite{poynting_pressure_1909}, a hanging piano wire under tension become longer when it is twisted (\textbf{Figure 1a}, left). The consequence is also that if the distance between the two ends is fixed, the torsion induces a stress normal to the shear plane (normal stresses) that tends to separate the two ends (\textbf{Figure 1b}, left) \cite{truesdell_mechanical_1952, billington_poynting_1986}. Developing normal stresses or axial deformations under torsion are two equivalent manifestations of the Poynting effect. Poynting found that the normal stress as a function of shear strain follows a quadratic relation with a positive coefficient \cite{poynting_pressure_1909, rivlin_torsion_1947, rivlin_large_1951},  now called the Poynting modulus. While conventional materials such as the piano wires of Poynting show a positive Poynting modulus (\textbf{Figures 1a} and \textbf{b}, left), complex materials such as biopolymer systems \cite{janmey_negative_2007, kang_nonlinear_2009, horgan_reverse_2015, mihai_positive_2011, de_cagny_porosity_2016, horgan_effect_2021} and designed structures \cite{misra_pantographic_2018} can exhibit a negative Poynting modulus, causing a shear-induced contraction under a fixed load (\textbf{Figure 1a}, right), or negative normal force at a fixed gap (\textbf{Figure 1b}, right). Designing metamaterials with exceptional mechanical properties originated from their structure rather than their composition has attracted much research in different disciplines of science \cite{bertoldi_flexible_2017, florijn_programmable_2014}. So far, mechanical metamaterials have been studied mostly under compression or tension. The response of metamaterials to direct shear \cite{liu_architected_2019}  and in particular, the Poynting effect have remained largely unexplored. Understanding the complex coupling between shear and normal responses in metamaterials provides insights for harnessing and programming their shear and Poynting moduli.

\par Pulling or pushing on isotropic linear elastic objects causes expansion or contraction, but torsion is not allowed \cite{timoshenko_theory_2010, coulais_as_2017}. In a limited number of recent studies, induced linear torsion by compression  have been uncovered in designed chiral structures  \cite{frenzel_three-dimensional_2017, shaw_computationally_2019, bessa_bayesian_2019, horgan_extension_2016, janbaz_ultra-programmable_2019} mainly due to broken spatial symmetry of the system. Hence we ask whether an object is capable of transforming pure compression into a nonlinear torsion, which we call the inverted Poynting effect (\textbf{Figure 1c}). The inverted Poynting effect is not a conventional material property and it has not been observed so far.  

\begin{figure}[t!]
	\centering
	\includegraphics[width=0.6\columnwidth]{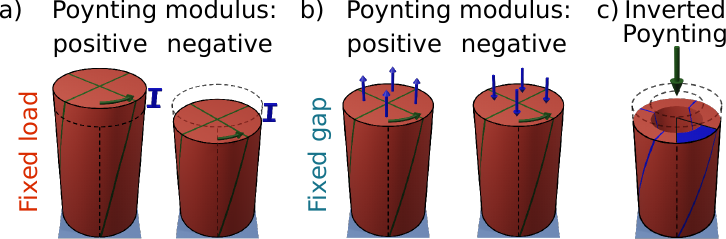}
	\caption{\label{fig1} Poynting effect. a) Applying torsion under fixed loads causes dilation in a material with a positive Poynting modulus (left) and contraction in a  material with a negative Poynting modulus (right). b) When the material is confined under a fixed gap, dilation and contraction will be manifested as positive (left) or negative (right) normal forces, respectively. c) Schematic of a cylindrical shell showing the inverted Poynting effect, where, in contrast to the Poynting effect’s manifestation in Poynting's original experiment, an applied compression induces a nonlinear torsion in the cylinder.}
\end{figure}

\par In this work, we aim to program the sign and magnitude of the Poynting modulus in a designed structure. We design a cylindrical metamaterial that exhibits a programmable Poynting modulus and an inverted Poynting effect, where compression induces a torsion.  The cylindrical metamaterial is composed of identical units (unit-cells)  consisting of beams with a suitably designed cross-section profile. By applying compression, the cylinder induces nonlinear and linear torsions. The sign and magnitude of the normal and shear responses of the designed structure are tuned by the interplay between buckling instability and self-contact interactions of the beams via a pre-compression applied prior to shear deformation. We present a simple spring model that reproduces our experimental results and characterizes the essential design parameters to obtain the inverted Poynting and to program the Poynting modulus. Our findings outline a strategy towards the rational designing of a programmable nonlinear elastic response of metamaterials with potential applications in engineering biomaterials functioning under torsional deformation and designing robot arms, soft rotational actuators and mechanical switches \cite{cianchetti_biomedical_2018, rus_design_2015, majidi_soft-matter_2019}. In soft robotics, for instance, coupling between torsion and compression  can be exploited for the design of twisters, rotational actuators, kinematic controllers and  pick and place end-effector \cite{schaffner_3d_2018, janbaz_ultra-programmable_2019}.

\par \textit{System and procedure:}
We design a hollow cylindrical shell composed of an array of unit-cells, which provides a network of nonuniform beams (\textbf{Figure 2a}) capable of side-buckling and self-contacting under compression \cite{overvelde_compaction_2012}. We apply compression and torsion deformations on the cylindrical metamaterial by clamping the structure between two custom-made plates of an Anton Paar rheometer and measure the mechanical responses. The experimental details are explained in the experimental procedure section.

\par We first conduct two series of compression experiments with different boundary conditions: in the first series the bottom side of the shell is clamped and the top side is free to rotate (‘torsion-free’), while in the second series rotation is not allowed at both sides (‘clamped’). Then we use the clamped boundary condition and perform two series of torsion experiments under fixed loads and fixed gaps.
\par The compression strain is defined as $\delta=|h-h_0|/h_0$, where $h_0=40.1 mm$ is the initial effective height of the cylindrical shell and  $h$ is its height after the pre-compression, under the compression force, $F$, with $\pm 0.2\%$ experimental error. Torsional angle, $\phi$, develops by applying shear force, $F_s$, and induces the axial deformation of $\delta_n$ under a fixed load or the normal force of $F_n$ under a fixed gap. In isotropic elastic materials, the shear stress, $\sigma_s$, is proportional to the shear strain, $\gamma$, $\sigma_s=G_s\gamma$, and the normal stress induced by shear follows a quadratic relation as a function of shear strain, $\sigma_n=G_n\gamma^2$ \cite{poynting_pressure_1909, rivlin_large_1951}. Here, we characterize the normal and shear responses of the structure with a Poynting modulus, $G_n$, and a shear modulus, $G_s$, respectively.  Normal and shear force responses of a cylindrical shell under torsion are given by $F_n=G_n J (\phi/h)^2$, and $F_s=\tau/R=G_s J \phi/Rh$, respectively, where $\tau$ is the torque around the axis of the shell, $R=(R_{max}+R_{min})/2$ is the mean radius, and $J=\frac{\pi}{2}(R_{max}^4-R_{min}^4)$ is the second moment of area of the shell.

\par \textit{Inverted Poynting effect and three regimes of structural rearrangements:}
Our designed cylindrical metamaterial is capable of showing the inverted Poynting effect by inducing torsion under compression (\textbf{Figures 2b} and \textbf{c}).  As shown in \textbf{Figure 2b} in the torsion-free compression, we observe the rotational buckling with an affine torsional deformation across the height of the sample. However, in the clamped compression experiment, the torsional deformation accumulated at the middle of the structure (\textbf{Figure 2c}). In both compression experiments, three distinct regimes of configurational changes are observable in the structure: (\romannumeral1) pre-buckling (\textbf{Figure 2a}), (\romannumeral2) buckling of the beams (\textbf{Figures 2b} and \textbf{c}, left), and (\romannumeral3) self-contact  (\textbf{Figures 2b} and \textbf{c}, right). \textbf{Figure 2d}, shows the compression stress, $\sigma$, rescaled by Young’s modulus of the elastomer, $Y=3.63 MPa$, as a function of compression strain, $\delta$, for the torsion-free (dashed curve)  and clamped (solid curve) experiments, respectively. The three regimes have different effective stiffnesses due to their structural configurations. The effective stiffness of the system in each regime, $Y_{eff}$, is determined by the slope of the curve times Young’s modulus of the elastomer. 

\begin{figure}[t!]
	\centering
	\includegraphics[width=0.5\columnwidth]{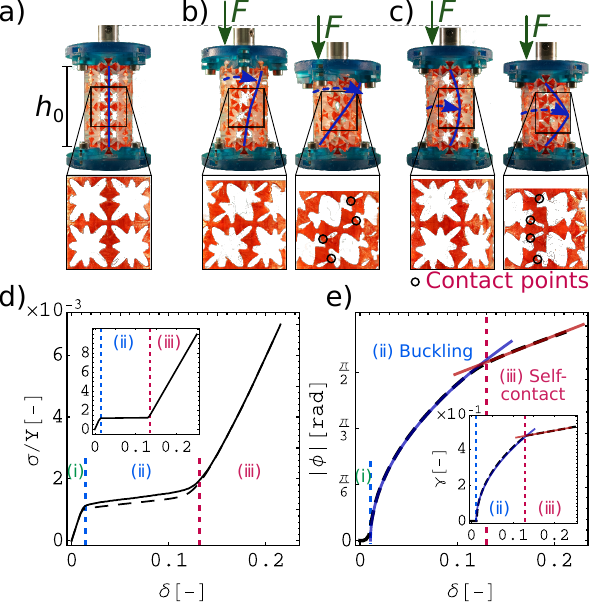}
	\caption{\label{fig2} Experimental setup and compression tests. a) 3D printed structure is clamped between two plates. Unit-cells and nonuniform beams are magnified.  b,c) The compressed cylinder at the buckling (left) and self-contact (right) regimes in torsion-free (b) and clamped (c) compression experiments. Deformed unit-cells are magnified. d) Rescaled nominal normal stress, $\sigma/Y$, as a function of compression strain, $\delta$, for torsion-free (dashed curve) and clamped (solid curve) experiments. e) Torsional angle, $\phi$, as a function of compression with square root (blue) and linear (red) fits, in the buckling and self-contact regimes, respectively. Insets are the predictions of the model and have the same axes and units as the main plots, except for the vertical axis in the inset of (e), which represents the shear strain in the 2D model, $\gamma$. $\gamma$ can be converted to the equivalent torsion in our cylindrical shell via $\phi=(h_0/R)\gamma$. Vertical dashed blue and red lines show the onset of buckling and self-contact regimes, respectively. Three regimes are marked with (i), (ii) and (iii).}
\end{figure}
In the pre-buckling regime ($\delta < \delta_b$, where $\delta_b=0.012$ is the compression strain for the onset of the buckling regime) the vertical beams are stable and resist buckling (\textbf{Figure 2a}), resulting in a relatively high stiffness, with $Y_{eff} =0.095Y$, in both compression tests. In the buckling regime ($\delta_b \le \delta < \delta_c$, where $\delta_c=0.13$ is the compression strain for the onset of the self-contact regime),  the stress remains almost constant and the structure softens ($Y_{eff}=0.004Y$), due to the buckling instability in the beams (\textbf{Figures 2b} and \textbf{c}, left). Finally, in the third regime ($\delta \ge \delta_c$), the structure becomes stiff again, with $Y_{eff}=0.06$, due to the self-contact between the beams (\textbf{Figures 2b} and \textbf{c}, right).
\par In the torsion-free experiments, compression induces shear deformation, with distinct behaviors at buckling and self-contact regimes (\textbf{Figure 2e}). The coupling between the compression and torsion, in the buckling regime, is nonlinear and the best fit to the experimental torsional angle as a function of compression gives $|\phi|=4.9\sqrt{\delta - \delta_b}$ (blue curve in \textbf{Figure 2e}). The emergence of this square root relation is due to the buckling of the internal beams \cite{coulais_discontinuous_2015}. However, in the self-contact regime, compression and torsion are linearly proportional  (red line in \textbf{Figure 2e}). Whereas linear coupling between compression and torsion has been achieved before for chiral structures \cite{frenzel_three-dimensional_2017, shaw_computationally_2019, bessa_bayesian_2019}, here for the first time nonlinear couplings between compression and torsion in an originally achiral structure have  been studied. Thus, the cylindrical metamaterial translates an axial compression to a nonlinear torsion (the inverted Poynting effect) or linear torsion depending on the amount of compression. In the following, we investigate the Poynting response of the clamped structure under different loading conditions.

\par \textit{ Poynting modulus under fixed loads/fixed gaps:}
To quantify both manifestations of the Poynting effect for our metamaterial, we apply torsional deformation under fixed loads (\textbf{Figure 1a}) or fixed gaps (\textbf{Figure 1b}) and follow its normal responses. The first loading scenario is equivalent to Poynting’s original experiment \cite{poynting_pressure_1909}. Initially, the clamped structure is loaded under a force $F$,  resulting in an axial compression strain, $\delta$. Then the torsion is applied on the top boundary while $F$ remains constant, causing an axial strain of  $\delta_n$. The axial strain and applied shear forces are shown as a function of torsion for a range of loads in \textbf{Figures 3a} and \textbf{b}, respectively. To quantify the second manifestation of the Poynting effect (torsion under fixed gap), we apply torsion on the pre-compressed shell while the height of the structure is fixed at $h$. Here, $F$ is the force needed for the pre-compression and $F_n$ is the torsion-induced normal force; $F+F_n$ is the total axial force response of the pre-compressed shell under torsion. \textbf{Figures 3c} and \textbf{d} show the normal forces and applied shear forces as a function of torsion under fixed gaps, respectively. For both series, the normal responses behave quadratically (\textbf{Figures 3a} and \textbf{c}) while the shear responses behave linearly (\textbf{Figures  3b} and \textbf{d}) as a function of torsion in the limit of small torsional deformation, $\phi <0.2 rad$. Initially, the non-compressed cylinder shows a contraction/negative normal force under torsion. While not a common material response, the negative Poynting modulus has been observed and investigated in biopolymer networks \cite{janmey_negative_2007, kang_nonlinear_2009, horgan_reverse_2015}. The origin of the negative Poynting modulus in biopolymers is rooted in the expulsion of water from their porous networks under deformation allowing them to shrink \cite{de_cagny_porosity_2016}. Similarly, in our metamaterial, the presence of voids allows to circumvent the volume conservation and to stretch the beams under torsion, which leads to negative normal responses. By applying a pre-compression on our metamaterial the curvatures of the normal response curves in both fixed load and fixed gap experiments (\textbf{Figures 3a} and \textbf{c}) change their sign and magnitude similarly. This indicates that the Poynting response of the structure is independent of whether the gap or the load was fixed, however, its sign and magnitude can be tuned by the level of pre-compression.

\par \textit{Programmable Poynting and shear moduli:}
To determine  how to program the nonlinear moduli of the metamaterial, we quantify shear and Poynting moduli as a function of pre-compression. For torsion under fixed gap experiments, the coefficients of the fits to the normal force, $F_n$, and shear force data, $F_s$, in \textbf{Figures 3c} and \textbf{d}, rescaled by $J/h^2$ and $J/Rh$, give the Poynting ($G_n$) and shear moduli ($G_s$), respectively. For torsion under fixed load experiments, we define the coefficient of the quadratic fits in \textbf{Figure 3a} rescaled by $J/A_sY_{eff}h^2$ as the Poynting modulus, where $A_s$ is the cross-section area. \textbf{Figures 3e} and \textbf{f} show the Poynting and shear moduli rescaled by Young’s modulus of the elastomer as a function of  compressive strain for both loading scenarios.

\begin{figure}[t!]
	\centering
	\includegraphics[width=0.5\columnwidth]{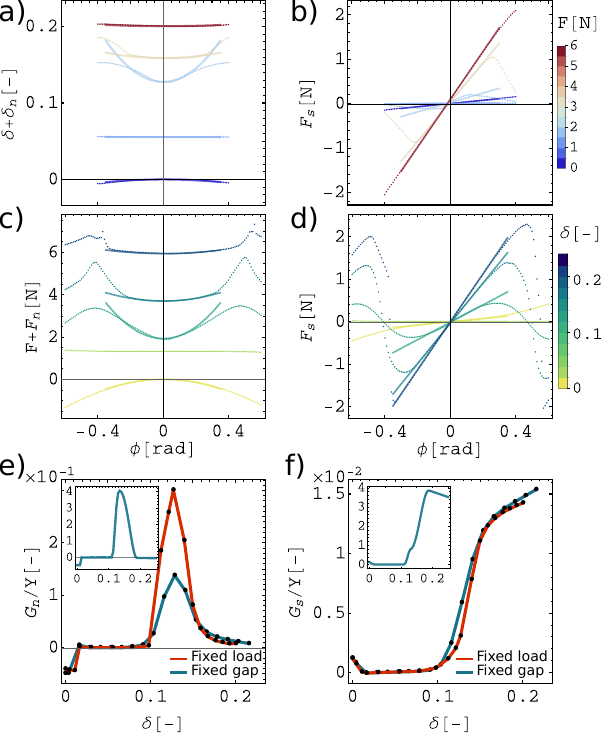}
	\caption{\label{fig3} Poynting and shear moduli. a,b) Fixed-load experiments: axial strain, $\delta+\delta_n$, (a) and shear force, $F_s$, (b)  as a function of torsional angle, $\phi$, while the structure is loaded under a fixed force ($F$, color scale). For the sake of clarity, only data for five experiments are shown.  c,d) Fixed-gap experiment: compression plus normal forces, $F+F_n$, (c)  and shear force, $F_s$, (d) as a function of $\phi$ at different levels of pre-compression strain ($\delta$, color scale). e,f) Poynting (e) and shear (f) moduli rescaled by Young’s modulus of the bulk, $Y$, as a function of pre-compression strain, $\delta$, for both loading scenarios calculated by fitting (solid curves) data points in a-d. Insets are the modeling results in the fixed gap boundary condition and have the same axes and units as the main plots.}
\end{figure}
The Poynting modulus in the pre-buckling regime is negative. For the intermediate pre-compressions (buckling regime), the Poynting modulus becomes zero. However, by approaching the self-contact regime, it rapidly increases and reaches a maximum value at $\delta \approx \delta_c$, where the transition from the buckling to the self-contact regime occurs. The obtained moduli from both loading scenarios coincide as expected, apart from a deviation occurring at this transient deformation. By further increasing the strain $G_n$ decays sharply and approaches zero. Both shear moduli calculated from the two loading scenarios coincide as well (\textbf{Figure 3f}). The shear modulus, $G_s$, first decreases to zero after the transition from the pre-buckling to the buckling regime. Subsequently, $G_s$ increases sharply at the onset of the self-contact regime and keeps increasing at higher pre-compressions but with a lower rate.  The absolute value of $G_n/G_s$ varies from 0 to about 13 for our metamaterial, whereas for a bulk cylindrical shell with the same dimensions and made of the same elastomer $G_n/G_s \approx 0.5$, in agreement with $G_n/G_s=5/8$, for an incompressible isotropic rubber \cite{rivlin_large_1951}. Thus, by tuning the level of pre-compression we can program the magnitude and sign of the Poynting modulus and even eliminate it, In addition, our structure shows a strong potential for tuning the shear modulus over a wide range.

\par \textit{Oscillatory Poynting modulus in large deformation:}
In \textbf{Figures 3a-d}, we observe deviations from the quadratic response with a nonmonotonic behavior at large torsions. To understand this behavior at large deformations, we follow the structural changes and mechanical responses under large torsional deformation at a fixed pre-compression in the self-contact regime ($\delta_c$). \textbf{Figure 4a} shows sequential images of the structural changes in our experiment. In \textbf{Figure 4b}, we observe a periodic oscillation in both normal (solid line) and shear (dashed line) forces as a function of torsion amplitude. In the course of torsion, one layer slides over another layer by snapping its beams from tilted to vertical and again tilted configurations, causing a local maximum in the normal force. Since the rearrangement occurs layer by layer, the number of the peaks of the normal response is set by the number of layers ($l=4$).  For large deformation experiments we determine the Poynting and shear moduli as  $G_n=(h^2/2J)(\partial^2F_n/\partial \phi^2)$, and $G_s=(Rh/J) (\partial F_s / \partial \phi)$. \textbf{Figure 4d} shows both rescaled $G_n$ (solid line, left axis) and $G_s$ (dashed line, right axis) oscillate between positive and negative values. Negative values of $G_s$ accompanied by negative slopes in the curves of shear force indicating the occurrence of the snap instability. Thus, snap instability and self-contact under large torsional deformation result in oscillatory nonlinear moduli that are not allowed in conventional materials and rare in metamaterials.
\begin{figure}[t!]
	\centering
	\includegraphics[width=0.5\columnwidth]{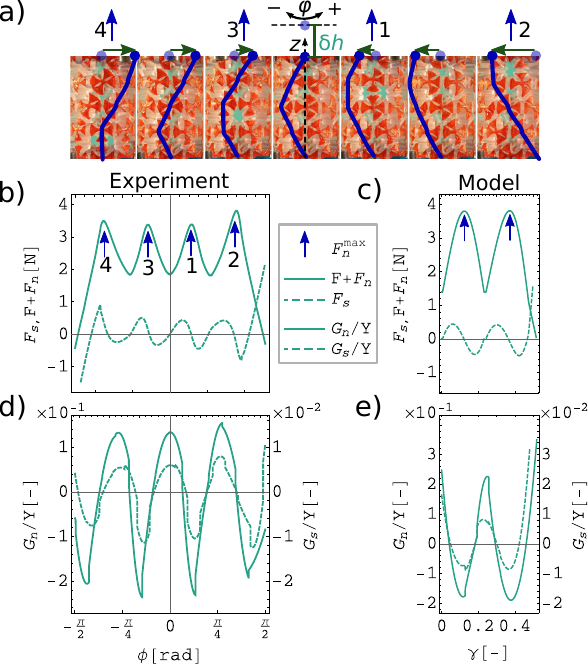}
	\caption{\label{fig4} Periodic Poynting moduli in large strain deformation  under the pre-compression of $\delta=0.13$ in the self-contact regime.  a) Sequence of images showing the internal deformations of the cylindrical structure during one cycle of a large amplitude torsion experiment. b,c) Normal response, $F+F_n$, (solid) and shear force, $F_s$, (dashed) as a function of torsion under large torsional deformations, for experiment (b) and model (c).  d,e) Rescaled Poynting, $G_n/Y$, (solid) and shear, $G_s/Y$, (dashed)  moduli as a function of torsion for experiment (d) and model(e). (c) and (e) are predictions of the model only for positive shear deformations ($\gamma \ge0$) with a fixed gap boundary. The results of negative shear are mirror images of these curves, with four peaks in one deformation cycle.}
\end{figure}
\par \textit{ Theoretical predictions vs. experimental results:}
We model the experimental system by using a 2D square network of  Hookean springs  and  energy minimization to predict the configuration  and mechanical responses of the system. Details of this model are explained in the modeling procedure section. The results predicted by the model successfully reproduce all the qualitative features of the experiments, as shown in the insets of \textbf{Figures 2d}, \textbf{2e}, \textbf{3e}, \textbf{3f}, \textbf{4c}, and \textbf{4e}, with minor quantitative differences.   For example, the model’s maximum values of $G_n$ and $G_s$ are higher than observed experimentally (\textbf{Figures 3e} and \textbf{f}). This can be attributed to neglection of  the thickness of the beams’ middle point in the model, which excludes the bending possibility at the middle of beams and  leads to a higher $G_n$ and $G_s$. Using this model, the  onset of  the self-contact regime can be predicted analytically by $\delta_c\approx(1-\cos(\pi/2-2\alpha))+\delta_b\approx 0.13$, in a good agreement with the experimental observations. Moreover, we analytically predict the square root coupling of compression-torsion observed in the inverted Poynting experiment as $\gamma=\sqrt{2(\delta-\delta_b)}$  (\textbf{Figure 2e}, inset). Since shear deformation represents itself as a torsion on the cylindrical shell, by converting shear strain to torsion using $\phi=(h_0/R)\gamma$, we predict a coefficient of $ 5.7 $ for the square root relation close to the experimental value of $ 4.9 $ (\textbf{Figure 2e}).
The simple linear spring model identifies buckling and self-contact as the minimum ingredients to achieve programmable/inverted Poynting effects and confirms the structural origin of the nonlinear responses.

\par In conclusion, we have engineered a cylindrical metamaterial with programmable Poynting and shear responses. We showed that our designed structure is capable of exhibiting the inverted Poynting effect by translating an axial compression to a nonlinear torsion, in contrast to conventional elastic materials. We also succeeded in programming the Poynting modulus by varying the level of pre-compression/loading prior to torsion.  We switched the sign of the Poynting modulus and tuned its value over a wide range, including even eliminating it. Furthermore, we successfully modeled and studied the system using an energy minimization method. The model identifies buckling of the ligaments and self-contact as the essential design elements to achieve programmable and inverted Poynting in a metamaterial. Our analytical approach opens avenues for bottom-up programming of the shear and normal mechanical responses of metamaterials based on self-contact as a mechanical feedback mechanism.  Besides the fundamental importance of understanding nonlinear shear and normal moduli, their  programmability provides a groundwork to numerous possible applications in solid mechanics. For instance, our system is capable of translating a unidirectional motion into torsion and of switching the mechanical forces. As these are relevant features of machines, we anticipate applications in designing machine materials, robot hands, mechanical force switches, and rotational actuators with with simpler and more efficient mechanism compared to the conventional mechanisms. Additionally, the ability to program  the Poynting effect inspires applications in biomechanics where couplings of torsional and axial deformations are ubiquitous (e.g.  tendons, cartilages, and cardiovascular systems) \cite{ taber_nonlinear_2004, criscione_mechanical_1999, horgan_finite_2012, horgan_extension_2016} and may lead to engineer novel biological tissues,  implants, and external prosthetic devices functioning under torsion and compression \cite{cianchetti_biomedical_2018, rus_design_2015, majidi_soft-matter_2019}. 

\par \textbf{Methods, materials, and modeling}
\par \textit{Experimental procedure:}
The cylindrical beam network is created by extracting an evenly distributed circular pattern of eight voids, with a 4-fold symmetry contour, in $l$ number of layers from a cylindrical shell with an outer radius of $R_{max}=12.5mm$, and an inner radius of  $R_{min}=7.5mm$. We use the polar function $s(u)=c [(1-(a+b))+a \cos (4 u)+b \cos (8 u)]$ to create shape of the pores with a 4-fold symmetry contour, where, $s(u)$ is the radius at the polar angle $u$, $x=s(u) \cos(u)$, and $ y=s(u) \sin(u) $. In this equation, $a$ and $b$ are the shape tuning parameters, and $c$ sets the pore's size. Considering $a=b=0$, we can create a circle with a radius of $c$. Overvelde et al. showed that a 2D network with the flowing parameters for the pore shape of the unit-cell, $a=-0.21$ and $b=0.28$ (\textbf{Figure 5a}), exhibits side-buckling under compression \cite{overvelde_compaction_2012}. This shape also provides a nonuniform profile for the cross-section of the beams that form the structure. We use the same parameters to create our unit-cells. First, in CAD software (Blender 2.8), we place the pore contour (\textbf{Figure 5a}) at the outer surface of our cylindrical shell so that the plane of the pore is parallel to the axes of the shell. Then, we extrude this pore shape in the radial direction towards the center of the cylinder and create the void volume (\textbf{Figure 5b}). Carving a radial array of $ 8 $ voids evenly distributed around $2\pi rad$, in $l$ layers leaves us with a cylindrical network of nonuniform beams composed of $8 \times l$ unit-cells. We set the pore size parameter as $c=4.7mm$, which gives the minimum beam thickness of $0.8mm$ at the shell's outer surface. We create extra half layers on top and bottom and, finally, clamp the structure with two O-rings with the same maximum and minimum radii as the shell and a height of $3mm$ (\textbf{Figure 5c}). These features enable us to clamp two sides of the cylindrical metamaterial during the measurements.
\begin{figure}[t!]
	\centering
	\includegraphics[width=0.5\columnwidth]{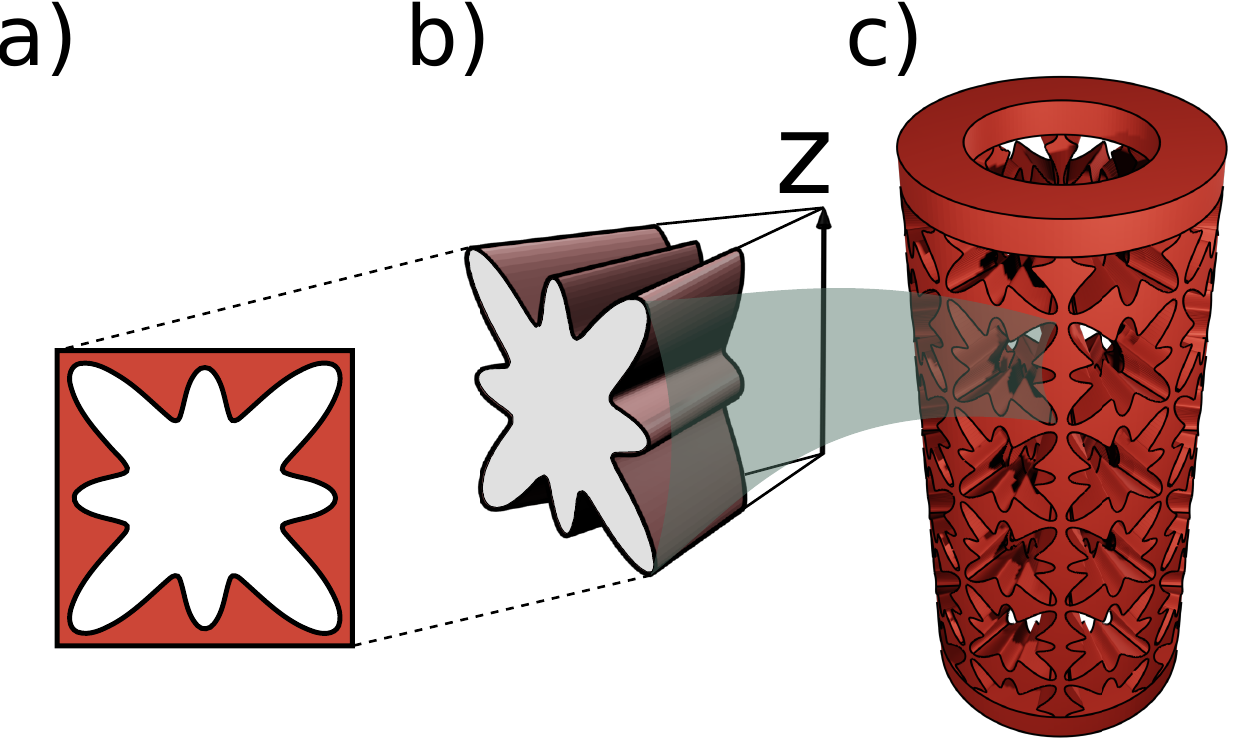}
	\caption{\label{fig5} a) The pore contour. b) CAD model of the void, created by extruding the pore contour in radial direction toward $z$ axis. c) An isometric view of the CAD model of the cylindrical metamaterial.}
\end{figure}
We 3D print the designed cylindrical metamaterial using a Formlab Form2 3D printer and elastic resin v1, with a resolution of $0.1 mm$. We conduct the compression and torsion experiments on our cylindrical metamaterial via an Anton Paar MCR302 rheometer with accuracies of $1\mu m$ (longitudinal displacements), $0.05 \mu rad$ (angular displacements), $0.005N$ (normal forces), and $1nNm$ (torques). Using a custom made plate-plate geometry, we clamp the upper and bottom sides of the shell (\textbf{Figure 2a}). The deformation angle, $\phi$, is positive when the torsion is clockwise. We 3D print a bulk cylindrical shell with the same elastic resin and dimensions as our cylindrical metamaterial, and perform a compression test to determine Young’s modulus of the bulk as $Y=3.63 MPa$.

\par \textit{Modeling:} 
We model the cylindrical metamaterial as a 2D square network of Hookean elastic beams with the length of $a_0$ that could either stretch or contract. The beams can bend at the connecting nods. Each beam has a pair of arc-shaped elastic arms, which are placed at the distance $r$ from each end of the beam and symmetrically spread by $\pm \alpha$ (\textbf{Figures 6a} and \textbf{b}). We study this model system in a 2-step deformation procedure: first, compression along the vertical axis, $z$ (\textbf{Figure 6c}), and then shear along the horizontal axis ($x$). The connecting arms to the beams are designed to mimic the beam profile's nonuniformity, and they can deliver self-contact under deformation. Self-contacts lead to elastic contractions of the arms that produce the reaction forces.
\par The total elastic energy due to deformation of each beam is composed of stretching energy, $ E_s = \frac{1}{2} k a_0^2 {e^2}$, and bending energy, $ E_b = k_b \theta^2 $. To calculate the contribution of self-contact in the elastic energy we assume that the arc-shaped arms have the same Hookean coefficient as the straight parts of the beam, $k$, and when subjected to a self-contact, their curvature remains constant but their length, $s$, changes through change of the arc angle, $ 2\alpha $; thus, $\delta s = 2 r \delta \alpha$. Since the deflection of the beam is divided between to contacting arcs, we can write $\delta \alpha = \delta \theta/2$, where $\delta \theta = \theta - \theta_c$ is the deflection of the beam after self-contact at $\theta_c = \frac{\pi}{2}-2\alpha$. Due to such deformation, the energy of each arm changes by $E_c=\frac{1}{2} k r^2 \delta \theta^2=\frac{1}{2} k_c \delta \theta^2$, which gives the elastic coefficient of the self-contact as $k_c=k r^2$. So the total energy for $l$ layers of the vertical beams is:
\begin{eqnarray}
E = && k_b \sum_{i=1}^{l}\theta_i^2 + \frac{1}{2} k a_0^2 \sum_{i=1}^{l}{e_i^2}+ k r^2 \sum_{i=1}^{l}{\delta \theta^2_i}
H\left[\delta \theta_i \right].
\end{eqnarray}
Since bending and self-contact occur symmetrically at both nodes of each vertical beam, their energy terms do not have the pre-factor $1/2$. Heaviside step function, $H$, represents the self-contact interactions; $ H[x] $ is $0$ for $x<0$ and $1$ for $x \ge 0$.

\begin{figure}[b!]
	\centering
	\includegraphics[width=0.5\columnwidth]{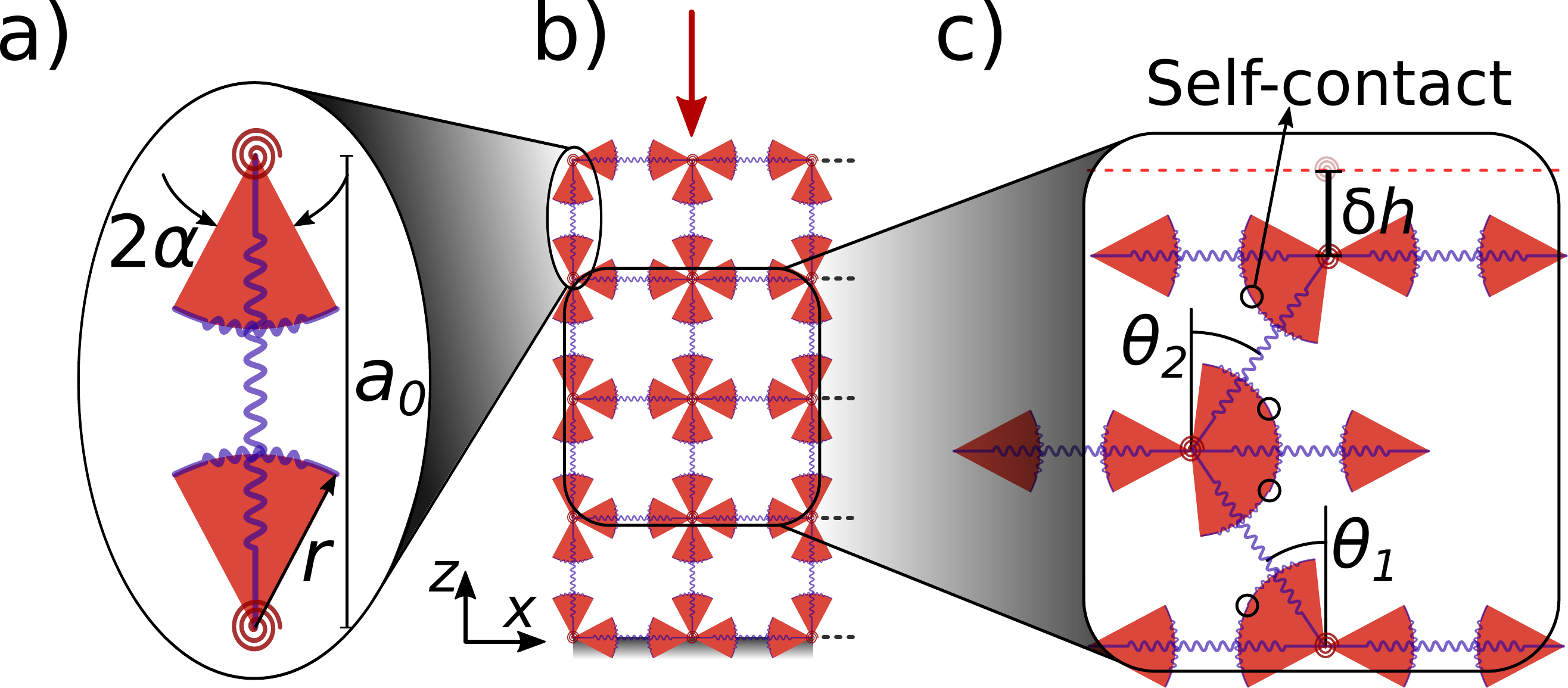}
	\caption{\label{fig6} a) Schematics of a non-uniform beam with associated parameters in our model. b) Schematic of the network of the beams in 4 layers. c) Under the pre-compression, beams are tilted by $\theta_i$ and when $\theta_i \ge \frac{\pi}{2}-2\alpha$ self-contact occurs.}
\end{figure}

By considering a periodic boundary condition in the horizontal direction, we can model a closed structure such as a cylinder. Since each horizontal beam is joined with two beams at its right and left sides, alteration of its orientation is not favorable. Under this condition and since we apply the compression and shear deformations uniformly at the top layer, horizontal beams remain horizontal. Thus we can infer one column structure's solution to the whole system and define the total energy of a system with $n$ unit-cells in each layer is $E_{tot}=nE$.
\par We use $n=8$, $l=4$, $\alpha=0.54rad$, and $r=4.7mm$, mimicking our experimental structure. We estimate the stretching coefficient from the force response in the compression experiment before buckling as $k=1331Nm^{-1}$. To estimate the bending stiffness, $k_b$, we consider a beam with a rectangular cross-section and use $k_b =M/\theta =YI/r_c\theta \approx (2YI/l_b) $, where $I$ is the moment of inertia, $l_b$ is the length of the beam, and $r_c$ is its bending curvature. This estimation predicts an order of magnitude of $ 10^{-4}Nm$ for the beams' bending stiffness. We finally calibrate the bending stiffness at $k_b=8.4\times 10^{-4}Nm$ to obtain the buckling at the same compression strain as in the experiment ($\delta_b=0.012$). We obtain the modeled system's configuration by numerically minimizing its energy (\textbf{Equation 1}) under different boundary conditions. Numerical minimizations are performed by Mathematica 11 using the Nelder-Mead method and a working precision of 15 digit numbers. We use the presented model to mimic the torsion-free compression, clamped compression, and fixed gap torsion experiments.


M. H. acknowledges funding from the Netherlands Organization for Scientific Research NWO, through NWO-VIDI grant No. 680-47-548/983. C.C. acknowledges funding from the European Research Council, through the Starting Grant No. 852587.

\vspace{12pt}
\textsuperscript{*} {aref.ghorbani@wur.nl}\\
\textsuperscript{$\dagger$} {mehdi.habibi@wur.nl}\\

\bibliographystyle{abbrv}
\bibliography{Poynting}

\end{document}